\nonstopmode
\documentclass[aip,pop,reprint,superscriptaddress]{revtex4-1}

\usepackage{graphicx}
\usepackage{epstopdf}
\usepackage{natbib}
\usepackage{amsmath,amssymb,amsfonts}
\usepackage[usenames,dvipsnames]{color}
\usepackage{subcaption}
\usepackage[normalem]{ulem}

\begin{document}

\title{NUMERICAL STUDY OF WEIBEL INSTABILITY DRIVEN BY ANISOTROPIC ELECTRON TEMPERATURE IN COLLISIONLESS PLASMAS}

\author{A.~Sladkov}\thanks{\emph{corresponding author e-mail:} \href{mailto:asladkov@ipfran.ru}{asladkov@ipfran.ru}}\affiliation{Institute of Applied Physics, 46 Ulyanov Street, 603950 Nizhny Novgorod, Russia}
\author{A.~Korzhimanov}\affiliation{Institute of Applied Physics, 46 Ulyanov Street, 603950 Nizhny Novgorod, Russia}

\begin{abstract}
We numerically investigate the process of generating magnetic fields from temperature anisotropy of electrons in collisionless initially uniform plasmas. We use a fully kinetic modeling and compare it against a hybrid modeling which treats ions kinetically and use ten-moment fluid model for electrons. The results of the one-to-one comparison show a good agreement in terms of the maximal magnitude of the self-generated magnetic field and similar trends during the non-linear stage of the instability. Additionally, we performed hybrid modelling of the instability without resolving electron spatial scales. In this case the results are only qualitatively the same however it shows that hydrodynamical approach can be used to some extent for the simulation of the Weibel instability in large-scale systems, including astrophysical environments and laser-produced plasmas.

\end{abstract}

\maketitle

\section{Introduction} \label{sec:intro}

The Weibel instability is a fundamental process in plasma physics, where it plays a crucial role in a variety of phenomena such as particle acceleration in collisionless shocks~\cite{spitkovsky2008}, gamma-ray bursts~\cite{medvedev1999}, magnetic field generation~\cite{widrow2012}, etc. It was first introduced by Weibel~\cite{weibel1959} as a mechanism of magnetic field generation in a collisionless plasma with anisotropic distribution of particle velocities and can be understood as a transformation of plasma free energy into the magnetic field energy due to an isotropization process. The nonlinear evolution of the instability is an object of great interest for more than fifty years in plasma physics. Already first numerical simulations supported the analytical result~\cite{morse1971} that initial anisotropy in the electron distribution function is resulting in growing of the current modes in range from $kc/\omega_p=0$ ($k$ is a mode wavenumber, $c$ is the speed of light and $\omega_p$ is the electron plasma frequency) to the $kc/\omega_p=\sqrt{A}$ where $A=T_{hot}/T_{cold}-1$ is an anisotropy parameter ($T_{cold}$ and $T_{hot}$ are respectively minimal and maximal mean squares of electron velocity distribution in different directions) and subsequent mode coalescing.

In literature we can find two main approaches to set initial anisotropy, either using two counterstreaming plasmas with beamlike distribution functions~\cite{califano2000,califano2001,silva2002} or as an uniform plasmas with usually a bi-Maxwellian velocity distribution as in the original numerical work~\cite{morse1971}. In order to distinguish between two branches, usually one adds term "thermal" to the Weibel instability~\cite{romanov2004} in the latter case. In laboratory plasmas, the Weibel instability has been observed in a variety of experimental setups, including interpenetrating plasmas generated by ns-scale laser pulses~\cite{fiuza2012,fox2013,huntington2015} or plasmas anisotropically heated by ps-scale pulses~\cite{quinn2012,gode2017,zhang2022}. Indirectly, the instability was studied during the laser ablation of ns-scale plasmas~\cite{fox2018, garasev2017}. The studies confirm the theoretical predictions~\cite{morse1971} and highlight the role of the electron temperature anisotropy in driving the instability. In the work, we focus on the thermal case using an uniform plasmas with an increased electron temperature in a certain direction. 

In the work we present a one-to-one comparison between fully kinetic results obtained with the particle-in-cell (PIC) description for both ions and electrons and hybrid-PIC with the ten moment closure for the electron fluid. The hydrodynamical description by the moments of distribution function~\cite{bychenkov1989,bychenkov1990,basu2002} has been widely used to study the Weibel instability, providing a computationally efficient method for the treatment of the nonlinear evolution. However, it is known to have limitations, as it cannot fully capture the small-scale electron structures. Despite these limitations, the hydrodynamical approach can provide good approximations of the anisotropy-driven Weibel instability growth rate and magnetic field evolution~\cite{romanov2004}. The convergence between the hydrodynamical and kinetic descriptions has been demonstrated in various works, the main goal of the work is to give insights on how to incorporate Weibel instability consequences in hybrid modeling.

This paper is organized into three sections: numerical model, results, and conclusions. In the numerical model section, we describe the details of our simulation setup, including the hydrodynamical equations we use, numerical methods employed, and initial and boundary conditions. We also discuss the parameters and assumptions used in our simulation. In the results section, we present the findings of our simulations by full-PIC and hybrid-PIC methods, comparing the evolution of the Weibel instability in the initially uniform unmagnetized plasma, the formation of current filaments and generation of the magnetic fields, and the effects on macroscopic plasma dynamics. Finally, in the conclusions section, we summarize our main findings and their implications for the understanding of the Weibel instability. We also discuss the limitations of our simulation and the potential for future research in this area. Overall, this paper aims to provide a first step in the study of the Weibel instability in laser-produced plasmas using full-scale hybrid simulations.

\section{Numerical model} \label{sec:mod}

For the reference modeling we use a fully kinetic PIC-code SMILIE\cite{derouillat2018}. Electrons and ions are described as particles with a mass ratio $1:10^{3}$. The simulation domain is a 2D rectangular box with sizes $L_X = 20c/\omega_{p}$, $L_Y = 20c/\omega_{p}$. As an initial condition, we use uniform unmagnetized plasmas with density 1 in the periodic box, ions are cold and electrons have an inplane temperature $T_{xx}=T_{yy}=$1keV and out-of-plane $T_{zz}=$10keV that gives the anisotropy parameter $A=T_{zz}/T_{xx}-1=9$. This choice is determined by the parameters of the laser plasmas we are going to study in future works. We use a 100$\times$100 grid corresponding to a mesh size equal to $\Delta x$=$\Delta y$=0.2$c/\omega_{p}$ in both directions, the time-step is $0.1\omega_p^{-1}$. The maximal wave number resolved by a such spatial resolution is $\pi/\Delta x<16$ which should be sufficiently enough for the chosen anisotropy. While ions are frequently considered immobile in studies~\cite{romanov2004,stockem2009} on the electron anisotropy-driven Weibel instability, in this work we take into account ion dynamics.

\begin{table}
\caption{Relations between normalization units for fully kinetic and hybrid modeling. For hybrid modelling two cases shown: in the middle -- the typical case when basic units are frequency and velocity and on the right -- the more suitable for our research case when basic units are density and pressure.}\label{tab:dim}
\begin{center}
\begin{tabular}{ccl}
\hline
\hline
PIC $(\omega_0)$ ~~ & HYBRID $(\Omega_0, V_0)$ & HYBRID $(n_0,P_0)$\\
\hline
$t^{PIC}_0$ = $\omega^{-1}_0$ & $t^{HYB}_0$ = $\Omega^{-1}_0$ & $t^{HYB}_0$ = $\frac{M_i}{e \sqrt{2 \mu_0 P_0}}$\\
$d^{PIC}_0$ = $\frac{c}{\omega_0}$ & $d^{HYB}_0 = \frac{V_0}{\Omega_0}$ & $d^{HYB}_0$ = $\sqrt{\frac{M_i}{ \mu_0 e^2 n_0 }}$\\
$B^{PIC}_0$ = $\frac{m_e}{e}\omega_0$ & $B^{HYB}_0 = \frac{M_i}{e}\Omega_0$ & $B^{HYB}_0$ = $\sqrt{2 \mu_0 P_0}$\\
$E^{PIC}_0$ = $\frac{m_e c}{e}\omega_0$ & $E^{HYB}_0 = \frac{M_iV_0}{e}\Omega_0$ & $E^{HYB}_0$ = $\frac{2 \sqrt{\mu_0} P_0}{\sqrt{M_i n_0}}$\\
$n^{PIC}_0$ = $\epsilon_0 \frac{m_e}{e^2} \omega^2_0$ & $n^{HYB}_0 =  \frac{M_i}{ \mu_0 e^2 V_0^2}\Omega^2_0$ & $n^{HYB}_0$ =  $n_0$\\
$P^{PIC}_0$ =  $\frac{m^2_e}{e^2 \mu_0} \omega^2_0$ & $P^{HYB}_0 =  \frac{M^2_i}{ e^2  \mu_0 }\Omega^2_0$ & $P^{HYB}_0$ =  $P_0$\\
$T^{PIC}_0$ =  $m_e c^2$ & $T^{HYB}_0 = M_i V_0^2$ & $T^{HYB}_0$ = $\frac{P_0}{ n_0 }$\\
\hline
\hline
\end{tabular}
\end{center}
\end{table}

For electron hydrodynamic modeling we use a hybrid code AKA~\cite{sladkov2020, akaGitHub}, built on classical principles~\cite{winske2003} of previous codes like
HECKLE~\cite{smets2011, smetsGitHub}, which keeps the ion description at the particle level and consider electrons as a massless neutralizing fluid described by a pressure tensor evolution equation~\cite{sladkov2021}. The solution of the evolution equation for the electron pressure tensor makes it possible to model the anisotropic temperature of electrons which is crucial for the description of the anisotropy-driven modes. For the cyclotron term integration in the pressure tensor evolution equation we use ion-to-electron mass ratio $\mu = M_i/m_e = 10^{3}$ and neglect the divergence of the electron heat flux.

The electromagnetic fields are treated in the low-frequency (Darwin) approximation, as we imply that the phase velocity of electromagnetic fluctuations is small compared to the speed of light. Neglecting the displacement current, we then write an electron Ohm's law~:

\begin{equation}
\mathbf E = -\mathbf V_i \times \mathbf B + \frac{1}{e n}(\mathbf J \times \mathbf B - \boldsymbol{\nabla} . \, \mathbf P_e) \label{ohm}
\end{equation}

In Eq.~(\ref{ohm}), $\mathbf E$ and $\mathbf B$ are electric and magnetic fields, $\mathbf V_i$ is the ion bulk velocity, $e$ is the elementary charge, $n$ is the electron density (equal to ions one), $\mathbf J$ is the total current density equal to the curl of $\mathbf B$, $\mathbf P_e$ is the electron pressure tensor and $\boldsymbol{\nabla} .$ is the divergence of the tensor. We use the explicit subcycling integration scheme for the six-component pressure tensor evolution equation~\cite{sladkov2021}. Electromagnetic fields are calculated on two staggered grids using a predictor-corrector scheme~\cite{winske1986}. The dynamics of the ions is solved using a first-order interpolation of the electromagnetic field~\cite{boris1972}.
 
For the hybrid modeling we use the same initial conditions as in fully kinetic case. To ease comparison we use the same values for density and pressure, in this case normalization units are related to each other as $n_0 = n^{PIC}_0$ and $P_0 = P^{PIC}_0$, then we have $t^{HYB}_0 = \frac{M_i}{\sqrt{2}m_e}t^{PIC}_0$, $d^{HYB}_0 = \sqrt{\frac{M_i}{m_e}}d^{PIC}_0$, $B^{HYB}_0 = \sqrt{2} B^{PIC}_0$, for the rest see Table~\ref{tab:dim}. For the chosen mass ratio for the  one-to-one comparison we use the spatial resolution $\Delta x=\Delta y=6\times 10^{-3}d_0$ and an appropriate timestep determined by the numerical scheme condition in this case is $10^{-5}\Omega_0^{-1}$.

\section{Results} \label{sec:res}

\begin{figure}
\includegraphics[width=.5\textwidth]{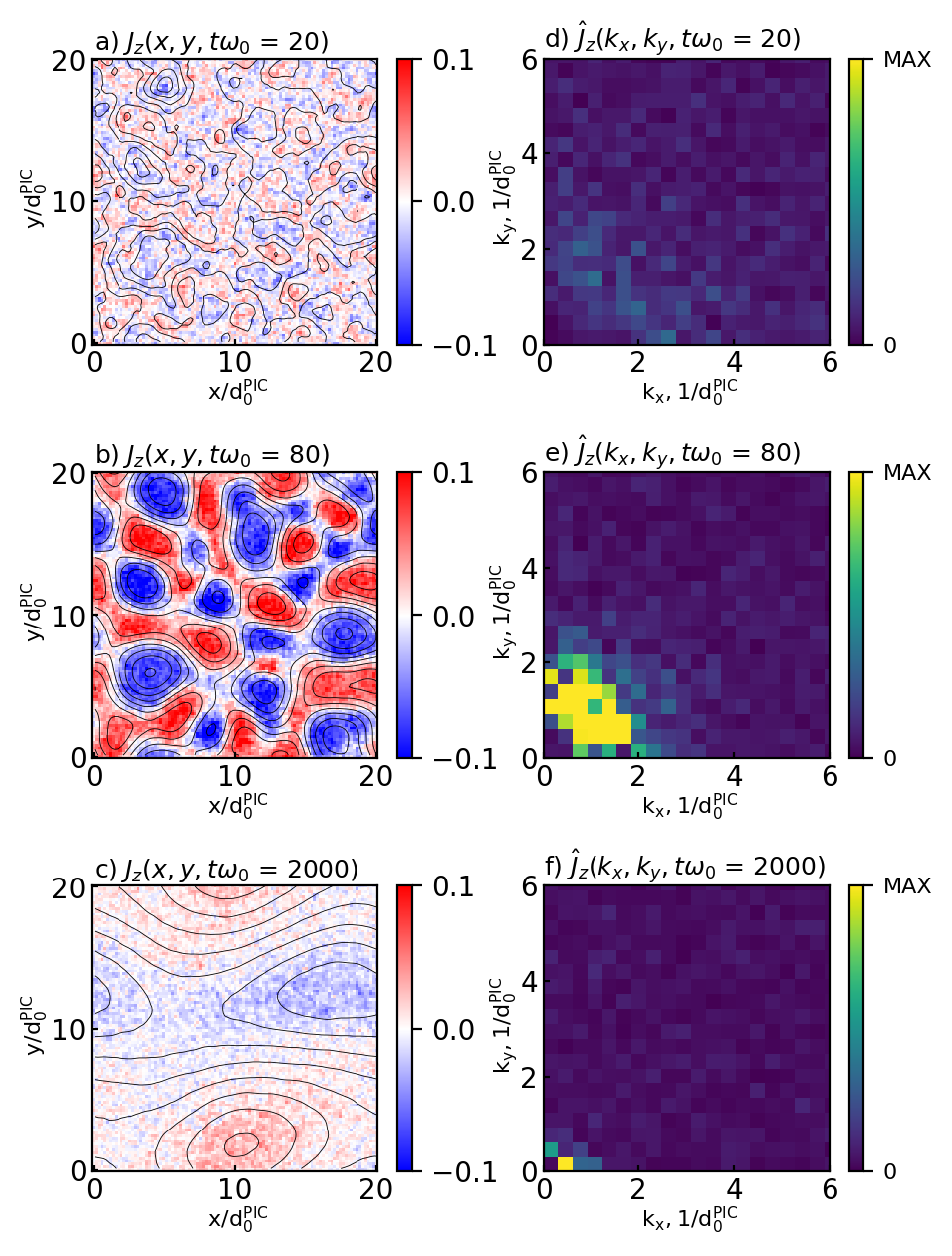}
\caption{Fully kinetic modeling. Color-coded out-of-plane current (a-c) and its spatial spectrum (d-f). Black lines represent inplane magnetic field lines. Currents are measured in $en_0^{PIC}c$ units.} \label{fig:pic2d}
\end{figure}

In this section we compare 2D simulations of the Weibel instability performed by the fully kinetic code SMILEI~\cite{derouillat2018} and a hybrid code AKA~\cite{akaGitHub}. Figure~\ref{fig:pic2d} shows snapshots from the fully kinetic modeling for the out-of-plane current (a-c) and its spatial spectrum (d-f). During the linear stage of the instability (a,d), the modes with $2 < kc/\omega_p < 3$ are dominating. During the nonlinear stage (b,e) the unstable modes excited in range $0<kc/\omega_p<\sqrt{T_{zz}/T_{xx}-1}$ are clearly seen which is consistent with the previous studies~\cite{morse1971,kocharovsky2016}. And in the end of the simulation (c,f) we can see only one long-wave mode survived with the wavenumber determined by the box size $kc/\omega_p \approx 0.1$.

\begin{figure}
\includegraphics[width=.5\textwidth]{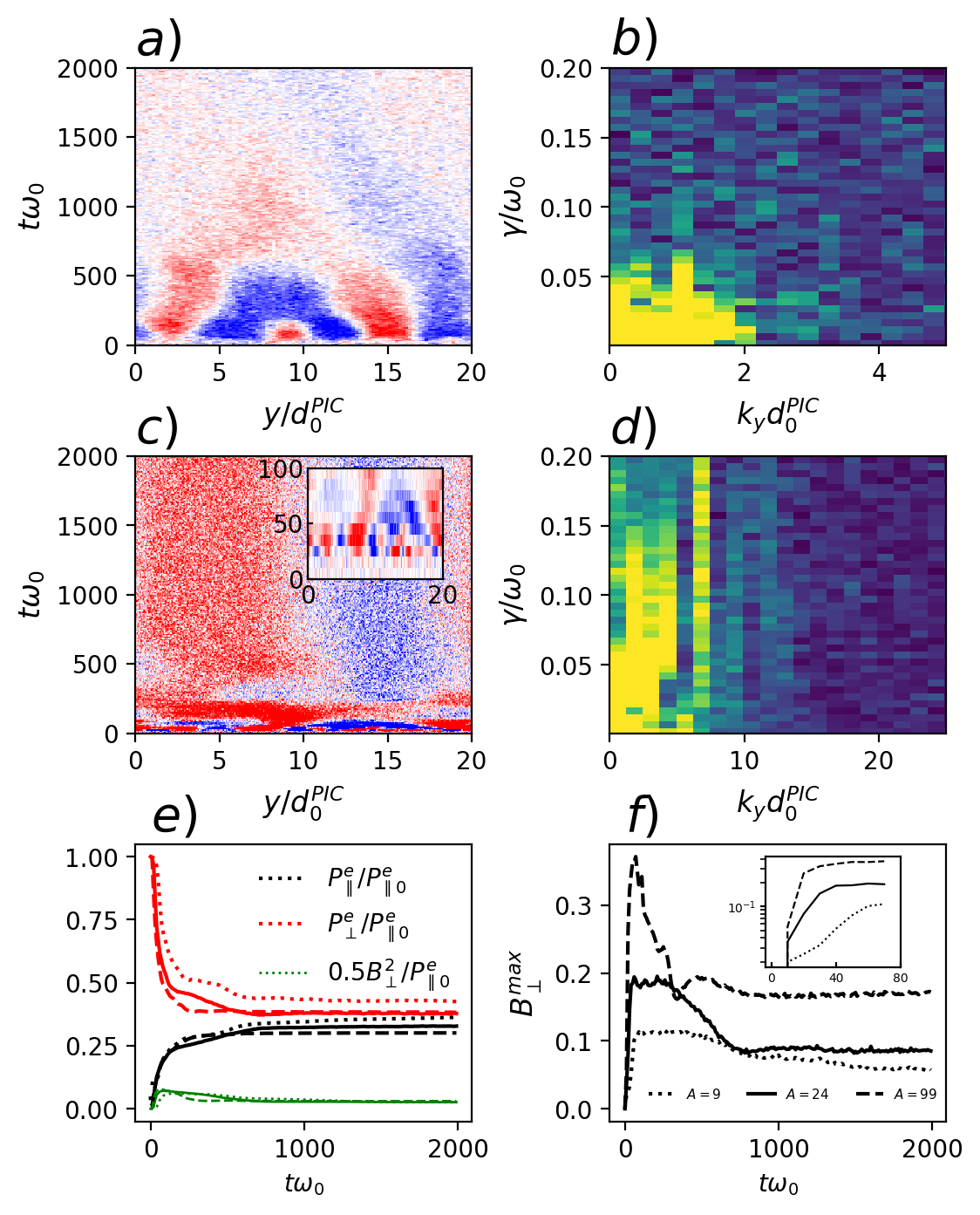}
\caption{Results of the fully kinetic modeling. Development of the out-of-plane current density (a,c) and its spectrum (b,d). Evolution of the box-averaged electron pressure tensor components (e) $P_{\parallel} = P_{zz}$ and $P_{\perp} = (P_{xx}+P_{yy})/2$. Evolution of the maximum of the self-generated magnetic field (f). Dashed curves are for A=99, full for A=24 and dotted for A=9. Panels (a,b) are for A=9 and (c,d) for A=99. Magnetic field is measured in $B_0^{PIC}$.}\label{fig:pic2d_spectrum}
\end{figure}

Figure~\ref{fig:pic2d_spectrum} shows an evolution in time for the 1D slice with color-coded out-of-plane current (a,c) and its spectrum (b,d), and an evolution of the box-averaged electron pressure (e) and in-plane magnetic field (f). The in-plane magnetic field (f) is exponentially growing for $t\omega_0<60$, then saturates for $60<t\omega_0<500$ and afterwards slowly decreases. Its maximal magnitude reaches $0.12B_0^{PIC}$, and in the end of the simulation it is kept constant around $0.06B_0^{PIC}$. During the quasistationary phase $80<t\omega_0<500$, the anisotropy level is slowly decreasing from $A(t\omega_0=80)\approx 2.6$ to $A(t\omega_0=500)\approx 0.5$. The residual anisotropy in the end of the simulation $A(t\omega_0=2000)\approx 0.2$. 

For the purpose of comparison we have performed two simulations with increased anisotropy parameters up to 24 ($T_{cold}$ and $T_{hot}$ are 1~keV and 25 keV) and 99 ($T_{cold}$ and $T_{hot}$ are 1~keV and 100~keV). The results are also shown at Figure~\ref{fig:pic2d_spectrum}. In the case $A=24$ the maximal magnitude of the magnetic field is $0.2 B_0^{PIC}$ which is consistent with the scaling $B\sim k \sqrt{P_{zz}-P_{xx}} \sim kA$. The saturated value in the end of the simulation is twice smaller than the maximal one. For $A=99$, the evolution happens faster (c), the spatial scale of the most energetic mode is broaden. The self-generated magnetic energy for three cases stays almost the same: $<7$~\% during the quasistationary stage and  $<3$~\% in the end of simulations. We should note that to resolve the spatial spectrum in the case with $A=99$ we used four times smaller cell size. Analysing the temperature evolution on panel (e) for three cases, we find that the maximal temperature is decreasing faster with increasing the anisotropy while the minimal temperature grows with the same rate in all the cases. The residual anisotropy stays within the range 0.3--0.4.

\begin{figure}
\includegraphics[width=.5\textwidth]{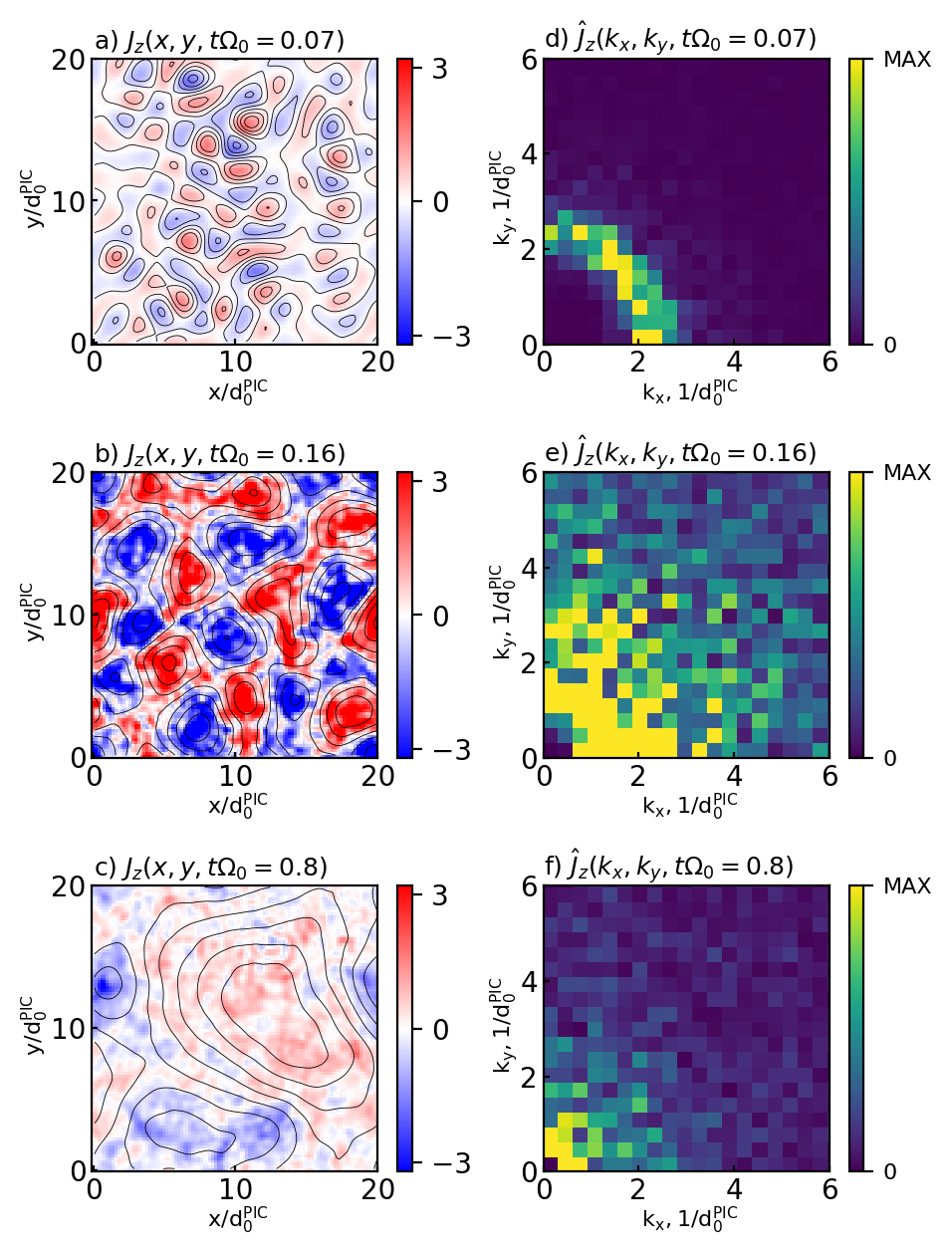}
\caption{Hybrid modeling. Color-coded out-of-plane current (a-c) and its spectrum (d-f). Black lines represent inplane magnetic field lines. Currents are measured in $en_0^{HYB}c\sqrt{m_e/2M_i}$ units.} \label{fig:hyb2d}
\end{figure}

\begin{figure}
\includegraphics[width=.5\textwidth]{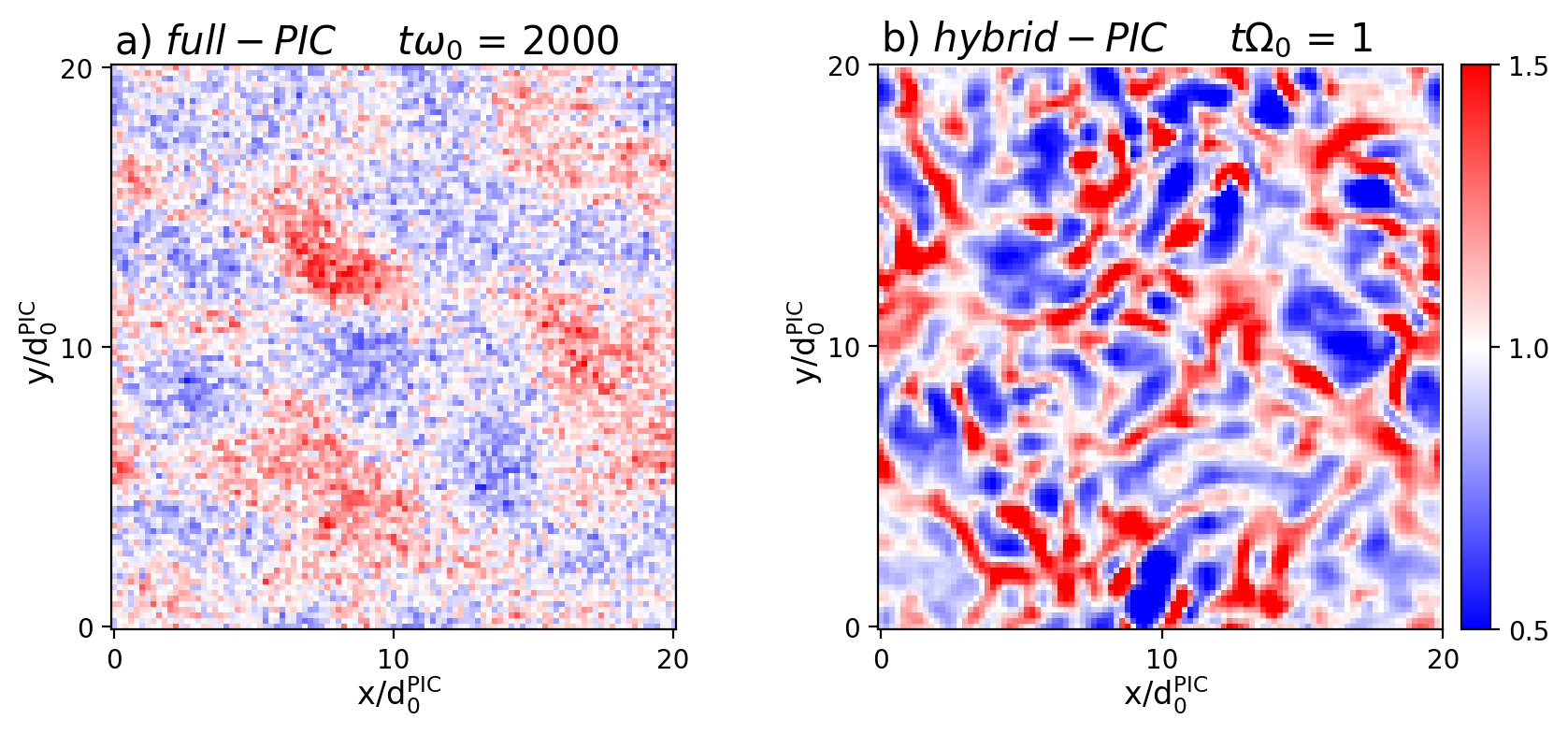}
\caption{Density for full-PIC (a) and hybrid-PIC (b) modeling.} \label{fig:dens}
\end{figure}

Figure~\ref{fig:hyb2d} displays snapshots from the hybrid modeling similar to the Figure~\ref{fig:pic2d}. As expected, during the linear stage (a,d) the unstable modes are close to the full-PIC result, the fastest growing mode has a wavenumber $2.1\omega_p/c$ while maximal value is limited by $3\omega_p/c$ as in the kinetic theory~\cite{morse1971} (see Appendix for details). While spatial distribution of currents and its spectrum during the nonlinear (b,e) and saturation (c,f) stages are comparable for both methods, in hybrid approach the coalescence rate turns to be three times higher and the growth rate of the magnetic field is ten times higher. In full-PIC $\gamma/\omega_0 \sim 0.02$ (which scales as square root from anisotropy level for the increased temperature cases), in hybrid-PIC $\gamma/\Omega_0 \sim 150$ what is in agreement with the linear analysis (see Appendix).

It is interesting to compare the density structures in the late times of the non-linear stage shown at Fig.~\ref{fig:dens}. We see density filaments aligned in the direction of the high electron temperature. The density variation can be about half of the initial value in hybrid modeling while it is $\sim 25$~\% in the full-PIC case, which highlights the importance of the ion mobility which was neglected in the previous kinetic studies~\cite{morse1971,romanov2004}. In the full-PIC case, at $t\omega_0=2000$ the density structures have smaller scales comparing to the current one, which is presumably due to ion inertia, as the coincidence of the density and current modes is achieved at much later times $\sim10^{4} \omega_0^{-1}$ (not shown).

\begin{figure}
\includegraphics[width=.5\textwidth]{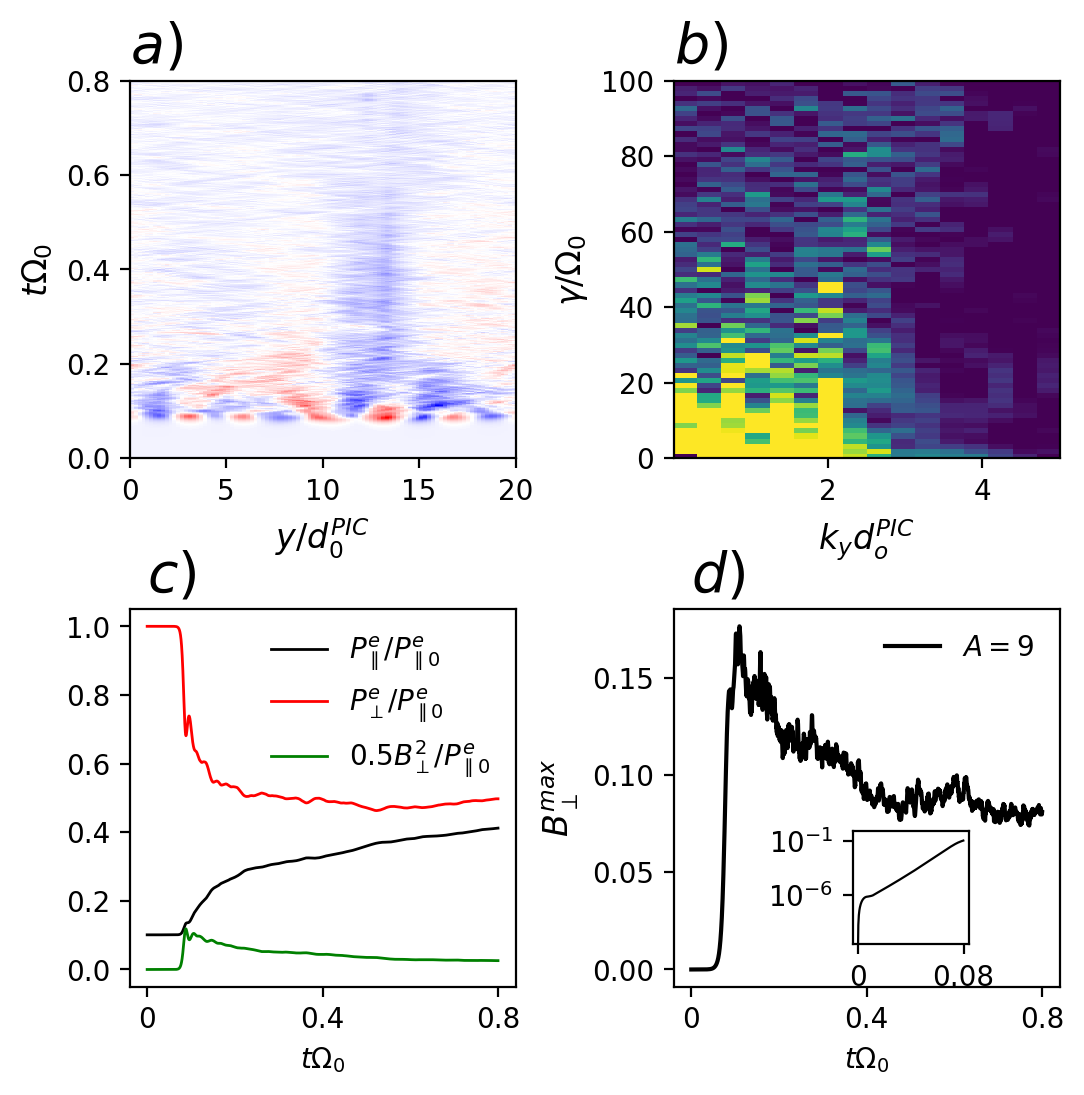}
\caption{Hybrid modeling for $A=9$ with full-PIC resolution $\Delta x$=$\Delta y=6 \times 10^{-3}d_0^{HYB}$, $\Delta t \Omega_0=2 \times 10^{-5}$.  Development of the out-of-plane current density (a) and its spectrum (b). Evolution of the box-averaged electron pressure tensor components (c) $P_{\parallel} = P_{zz}$ and $P_{\perp} = (P_{xx}+P_{yy})/2$. Evolution of the maximum of the self-generated magnetic field (d). Magnetic field is measured in $B_0^{HYB}$.} \label{fig:hyb2d_spectrum}
\end{figure}

Figure~\ref{fig:hyb2d_spectrum} shows hybrid modeling result similar to Fig.~\ref{fig:pic2d_spectrum}. It can be seen that the general behaviour is the same as in the full-PIC case. The residual anisotropy (panel c) stays around $A=0.3$, the maximal magnetic field magnitude (panel d) reaches $\approx 0.15B_0^{HYB}\approx 0.2B_0^{PIC}$ which is twice as much as in the full-PIC case. The self-generated magnetic energy stays $<12$~\% during the nonlinear stage and $<3$~\% in the end of simulations which is also comparable with the fully kinetic result. Important to note, that we cannot model significantly latter times neglecting the isotropization effect of the heat flux in the pressure tensor evolution equation as we observe a substantial numerical heating for an extremely small spatial resolution we use. The reason for this heating is that the driver terms being proportional to the electron velocity gradients $-2P_{ii}\partial_iu_i$ pump the diagonal components $P_{xx}$ and $P_{yy}$~\cite{sladkov2021}. Nevertheless, further research on the stability is required as well as on the electron heat flux closure. 

\begin{figure}
\includegraphics[width=.5\textwidth]{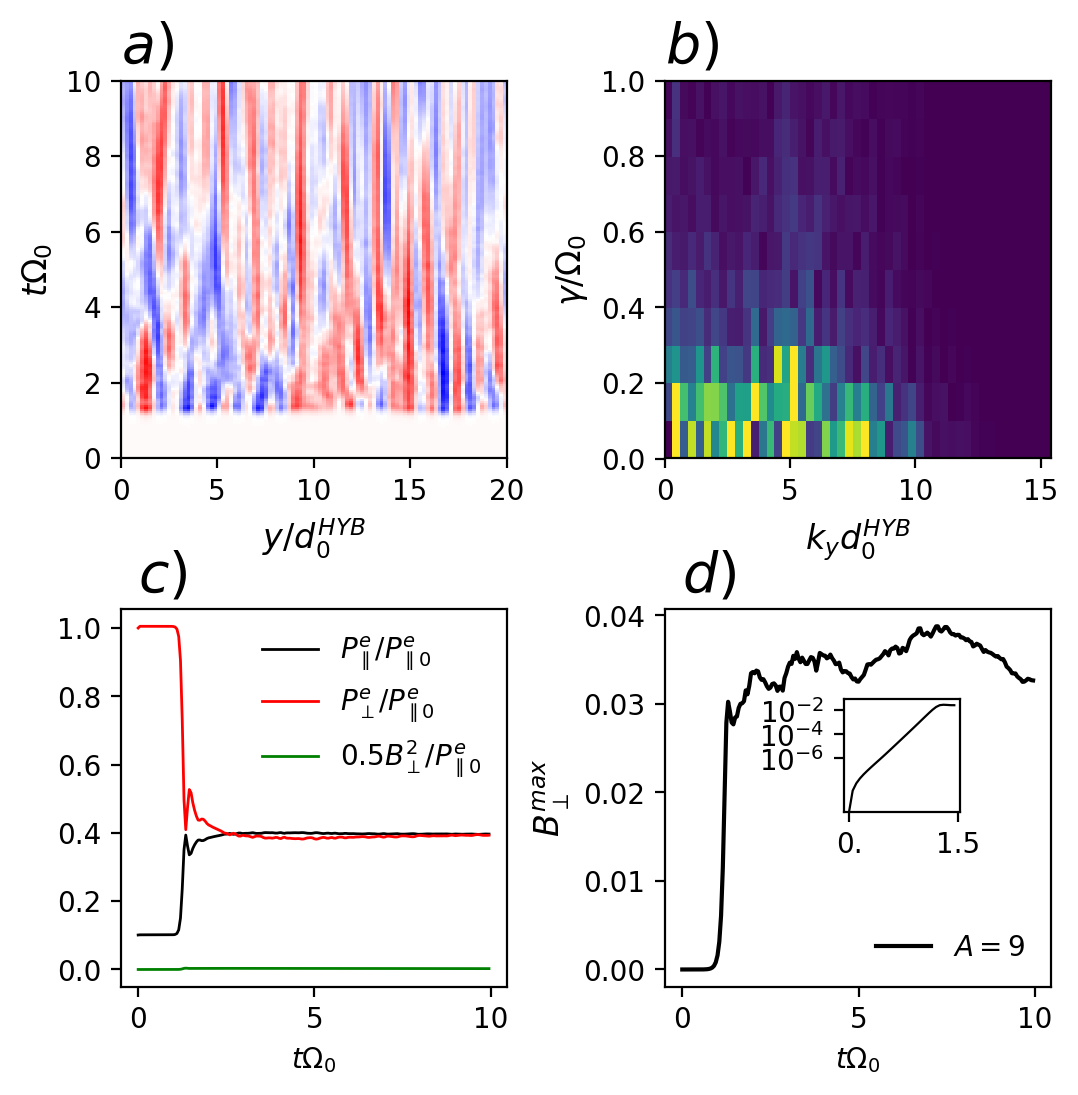}
\caption{Hybrid modeling for $A=9$ and ion-scale resolution $\Delta x$=$\Delta y$=0.2$d_0^{HYB}$, $\Delta t \Omega_0=5\times 10^{-3}$.  Development of the out-of-plane current density (a) and its spectrum (b). Evolution of the box-averaged electron pressure tensor components (c) $P_{\parallel} = P_{zz}$ and $P_{\perp} = (P_{xx}+P_{yy})/2$. Evolution of the maximum of the self-generated magnetic field (d). Magnetic field is measured in $B_0^{HYB}$.} \label{fig:hyb2d_res}
\end{figure}

The main goal of our work is to understand the limitations of the hybrid-PIC approach to the simulation of the Weibel instability in the case typical for that kind of modeling when the electron spatial scales are not resolved. So far, to make the one-to-one comparison we used high resolution which is two orders higher than the typical one using in hybrid modelling. Figure~\ref{fig:hyb2d_res} shows hybrid modeling result in a greater box with the ion-scale resolution $\Delta x=\Delta y=0.2 d_0^{HYB}$, $\Delta t=5\times 10^{-3}\Omega_0^{-1}$ similar to the results shown at Fig.~\ref{fig:hyb2d_spectrum}. We may note that the growth of the instability starts from the grid scale and the width of perturbations stays about few cells on sub-ion scale. In this case, the residual level of anisotropy is zero. Finally, the growth rate of the Weibel instability is 10 times slower than for the case of the electron-scale resolution which is expected as the low resolution effectively damps the fastest growing small-scale modes. The saturated magnitude of the generated magnetic field is two times smaller than for the appropriate resolution which is not dramatic but we do not resolve the peak of the magnetic filed with a such rough timestep.

In conclusion of the section, the anisotropy-driven Weibel instability is a fascinating plasma phenomenon that arises due to a difference in the temperatures of the plasma particles in different directions. Through the Weibel mechanism, this anisotropy leads to the growth of small-scale magnetic fields. The hydrodynamical description for electrons allows us to reproduce the spatial scales and its spectrum during the non-linear stage of the instability. However, it does not accurately reproduce the timescales involved. For ion-scale resolutions in hybrid modeling we have shown that the growth rate of the instability is significantly slower, while the spatial scales on later times are reproduced, and the residual magnetic fields are of the same order as in fully kinetic modeling. Further research in this area could lead to a deeper understanding of these discrepancies and their implications.

\section{Conclusions} \label{sec:con}

In this study, we compared the results obtained from a fully kinetic particle-in-cell modeling and a hybrid approach that treats ions kinetically and uses a ten-moment model for the fluid electrons. The comparison was made in terms of the spectrum, maximal magnetic field, and residual anisotropy level during the non-linear stage of the Weibel instability driven by the electron temperature anisotropy. The results show a good agreement between the two approaches, with similar trends observed in the behavior of the system. Specifically, the maximal magnitude of the self-generated magnetic field and the spectrum of the instability were found to be similar in both the full-PIC and hybrid-PIC models. Additionally, the residual anisotropy level was found to be similar in both models, indicating that the anisotropy level remains significant even during the non-linear stage of the instability. These findings suggest that the hybrid modeling approach may be a suitable alternative to fully kinetic modeling for studying Weibel instability driven by anisotropic electron temperature in collisionless plasmas.

The ability to incorporate the anisotropy-driven Weibel instability in hybrid codes with a rough ion-scale resolution has significant implications in the study of plasma physics. The Weibel instability is a fundamental mechanism responsible for the generation of magnetic fields in collisionless plasmas. However, accurately simulating the Weibel instability in fully kinetic codes can be computationally expensive, as it requires resolving both the electron and ion scales. With the ability to incorporate this instability in a hybrid code, researchers can simulate this phenomenon more efficiently while still capturing the essential physics. This will allow for more extensive studies of the Weibel instability's effects on plasma dynamics and could have important implications for fields such as laboratory astrophysics and fusion energy. 

In future studies, the hybrid modeling approach approved in this work may be useful for studying the Weibel instability in laser-produced plasmas. The hybrid approach offers a computationally efficient alternative to fully kinetic PIC simulations, making it feasible to simulate larger systems on longer time scales. Overall, we believe that the hybrid modeling approach has the potential to provide valuable insights into the dynamics of Weibel instability in laser-produced plasmas and could lead to new discoveries and applications in the field of plasma physics.

\section{Acknowledgments}
The research was supported by the Russian Science Foundation, project no. 19-72-10111. The authors would like to thank V.V. Kocharovsky, M.A. Garasev, and A.A. Nechaev for fruitful discussions. The
 simulations were performed on resources provided by the Joint Supercomputer Center of the Russian Academy of Sciences.

\section{Data Availability}
The data that support the findings of this study are available from the corresponding author upon reasonable request.

\section{Appendix} \label{sec:eq}

The system of equations for the linear analysis of the case with the increased out-of-plane electron temperature which generates inplane magnetic fields:
\begin{eqnarray}
\begin{cases}
\partial_x B_y - \partial_y B_x  = -\mu_0 env_z \\
\partial_t B_{x}  =  - \partial_y E_{z} \\
\partial_t B_{y}  =    \partial_x E_{z}\\
enE_z =  -\partial_x P_{xz} - \partial_y P_{yz} \\
\partial_t P_{xz}  =  -P_{xx}\partial_x v_z  - \frac{e}{m_e}B_y (P_{xx}-P_{zz}) \\
\partial_t P_{yz}  =  -P_{yy}\partial_y v_z - \frac{e}{m_e}B_x (P_{zz}-P_{yy})\\
\partial_t P_{xx}  =  2 \frac{e}{m_e}B_y P_{xz}\\
\partial_t P_{yy}  =  - 2 \frac{e}{m_e}B_x P_{yz}\\
\partial_t P_{zz}  =  -2(P_{xz}\partial_x v_z+P_{yz}\partial_y v_z) - 2  \frac{e}{m_e}(P_{xz}B_y-P_{yz}B_x)
\end{cases}
\end{eqnarray}

here we neglect advection term and divergence of the electron velocity in the pressure tensor evolution equation, both terms play nominal role for the field generation but it is important to keep both in numerical simulations for the purpose of stability on the long time scales. In the Ohm's law we keep only divergence of the electron pressure tensor.

The resulted system of equations for the inplane magnetic fields:
\begin{eqnarray*}
\begin{cases}
en\partial^2_{tt} B_{x}  =   \partial^2_{xy} (P_{xx}(\partial^2_{xx} B_y - \partial^2_{xy} B_x)/(\mu_0en)\\  - \frac{e}{m_e}B_y (P_{xx}-P_{zz}))\\ + \partial^2_{yy} (P_{yy}(\partial^2_{xy} B_y - \partial^2_{yy} B_x)/(\mu_0en) - \frac{e}{m_e}B_x (P_{zz}-P_{yy})) \\
en\partial^2_{tt} B_{y}  =    - \partial^2_{xx} (P_{xx}(\partial^2_{xx} B_y - \partial^2_{xy} B_x)/(\mu_0en)\\  - \frac{e}{m_e}B_y (P_{xx}-P_{zz}))\\ - \partial^2_{xy} (P_{yy}(\partial^2_{xy} B_y - \partial^2_{yy} B_x)/(\mu_0en) - \frac{e}{m_e}B_x (P_{zz}-P_{yy}))
\end{cases}
\end{eqnarray*}

after linearization we obtain the dispersion relation:
\begin{eqnarray*}
(-\omega^2 \mu_0e^2n^2 + P_{xx}k^2_xk^2_y + P_{yy}k_y^4  - \frac{\mu_0e^2n}{m_e} (P_{zz}-P_{yy})k_y^2)*\\(-\omega^2 \mu_0e^2n^2 + P_{yy}k^2_xk^2_y + P_{xx}k_x^4 + \frac{\mu_0e^2n}{m_e}(P_{xx}-P_{zz})k_x^2 ) \\ - (-P_{xx}k_x^3k_y - \frac{\mu_0e^2n}{m_e}(P_{xx}-P_{zz})k_xk_y - P_{yy}k_xk^3_y)*\\(- P_{xx}k^3_xk_y + \frac{\mu_0e^2n}{m_e} (P_{zz}-P_{yy})k_xk_y - P_{yy}k_y^3k_x)  = 0
\end{eqnarray*}

we use the following normalization
\begin{eqnarray}
\omega^{-2}_p & = & {\frac{m_e\epsilon_0}{e^2n}} ={\frac{m_e}{\mu_0c^2e^2n}} \nonumber\\
v^2_{Tx} & = & {\frac{P_{xx}}{nm_e}} \nonumber\\
v^2_{Ty} & = & {\frac{P_{yy}}{nm_e}} \nonumber\\
v^2_{Tz} & = & {\frac{P_{zz}}{nm_e}} \nonumber\\
A^{xz} &=& \frac{P_{xx}}{P_{zz}} \nonumber\\
A^{yz} &=& \frac{P_{yy}}{P_{zz}} \nonumber\\
A_{x} &=& 1-A^{xz} \nonumber\\
A_{y} &=& 1-A^{yz} \nonumber\\
\tilde\omega &=& \frac{\omega}{\omega_p} \nonumber\\
\tilde{\mathbf k} &=& \frac{c\mathbf k}{\omega_p} \nonumber\\
u_x &=& \frac{v_{Tx}}{c} \nonumber \\
u_y &=& \frac{v_{Ty}}{c} \nonumber \\
u_z &=& \frac{v_{Tz}}{c} \nonumber
\end{eqnarray}

finally we get the equation (tildes are omitted):
\begin{eqnarray*}
\omega^4 - \omega^2( u^2_{y}k^2_xk^2_y + u^2_{x}k_x^4 - u^2_{z}A_{x}k_x^2  + u^2_{x}k^2_xk^2_y + u^2_{y}k_y^4 - u^2_{z} A_{y}k_y^2)\\+ (u^2_{y}k^2_xk^2_y + u^2_{x}k_x^4 - u^2_{z}A_{x}k_x^2)*(u^2_{x}k^2_xk^2_y + u^2_{y}k_y^4 - u^2_{z} A_{y}k_y^2) \\ - (u^2_{x}k_x^3k_y - u^2_{z}A_{x}k_xk_y + u^2_{y}k_xk^3_y)*\\(u^2_{x}k^3_xk_y - u^2_{z}A_{y}k_xk_y + u^2_{y}k_y^3k_x)  = 0
\end{eqnarray*}

which has two solutions, the solution with minus sign does not include anisotropy dependencies while another one gives the following result:
\begin{eqnarray*}
\omega = \sqrt{k^2_x(u^2_{x}k_x^2 - A_{x}u^2_{z}) + (u^2_{x}+u^2_{y})k^2_xk^2_y + k^2_y(u^2_{y}k^2_y - A_{y}u^2_{z})}
\end{eqnarray*}

\begin{figure}
\includegraphics[width=.5\textwidth]{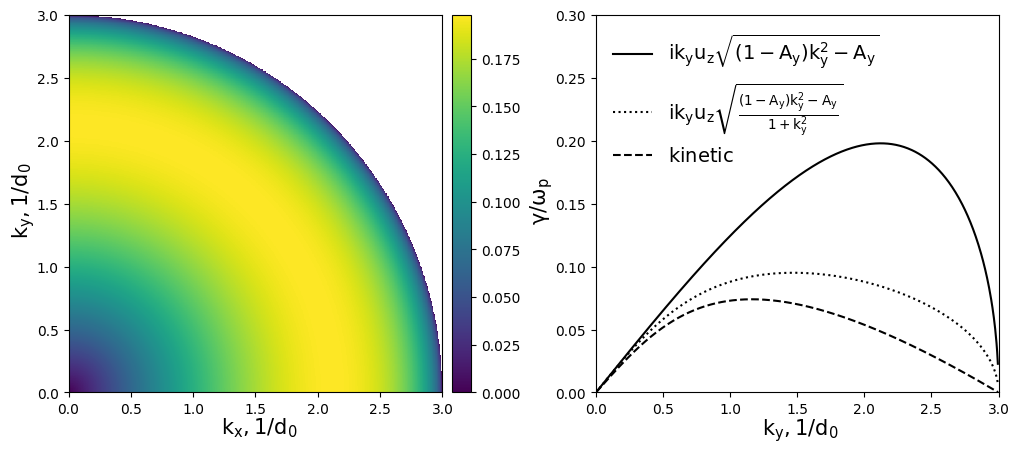}
\caption{Left panel - $\gamma(k_x,k_y)$, right panel - $\gamma(k_x)$ for $A_x=A_y=0.9$, $u_x=u_y=0.044$, $u_z=10u_x$. } \label{fig:analyt}
\end{figure}

the dependency for the fastest growing mode wave number can be easily obtained for a 1D case ($k_x = 0$)

\begin{eqnarray}
\omega= k_yu_{z}\sqrt{(1-A_y)k^2_y - A_{y}}
\end{eqnarray}

the maximal excited wave number is
\begin{equation}
    k_y^{max} =\sqrt{\frac{A_y}{1-A_y}} = \sqrt{\frac{P_{zz}}{P_{yy}}-1}
\end{equation}
what coincides with the kinetic~\cite{morse1971} and hydrodynamic~\cite{romanov2004} results.

The fastest growing mode has the wave number:
\begin{equation}
    k_y^{fast} =\sqrt{\frac{A_y}{2(1-A_y)}}=k_y^{max}/\sqrt{2}
\end{equation}

In order to get a known hydrodynamical solution~\cite{romanov2004}, we should take into account the electron inertia term in the Ohm's law being proportional to the time derivative of the electron velocity which gives the denominator $\sim \sqrt{1+k^2}$ shifting the maximum to the long waves (Fig.~\ref{fig:analyt}).

\bibliographystyle{apsrev}
\bibliography{biblio.bib}

\end{document}